\documentclass[runningheads]{llncs}
\usepackage[T1]{fontenc}
\usepackage{cite}
\usepackage{amsmath,amssymb,amsfonts}
\usepackage{algorithmic}
\usepackage{graphicx}
\usepackage{textcomp}
\usepackage{xcolor}

\usepackage{xcolor}
\usepackage{color}
\definecolor{gray75}{gray}{.25}
\definecolor{gray70}{gray}{.3}
\definecolor{gray60}{gray}{.4}
\definecolor{gray50}{gray}{.5}
\definecolor{gray40}{gray}{.6}
\definecolor{gray30}{gray}{.7}
\definecolor{gray20}{gray}{.8}
\definecolor{gray15}{gray}{.85}
\definecolor{gray10}{gray}{.9}
\definecolor{gray05}{gray}{.95}

\definecolor{graytable}{gray}{.95}

\definecolor{headertable}{rgb}{0.0, 0.0, 0.0}

\definecolor{steBoxLine}{rgb}{0.0, 0.0, 0.0}

\usepackage{comment}
\usepackage{hyperref}

\usepackage{booktabs}
\usepackage{multirow}
\usepackage{wrapfig}

\usepackage{url}

\usepackage{balance}

\usepackage{enumitem}

\usepackage{fontawesome}

\usepackage{cleveref}

\usepackage[colorinlistoftodos]{todonotes}

\usepackage{tabularx}
\usepackage{siunitx}
\usepackage{multirow}
\usepackage{colortbl}
\usepackage{longtable}
\usepackage{tabularray}
\usepackage{tabu}

\usepackage{dcolumn}

\usepackage{array}

\newcolumntype{P}[1]{>{\centering\arraybackslash}p{#1}}
\newcolumntype{a}{>{\columncolor{headertable}}l}
\newcolumntype{d}[1]{D{.}{.}{#1}}

\usepackage[]{mdframed}

\mdfdefinestyle{stebox4}{%
linecolor=steBoxLine,linewidth=2pt,%
leftmargin=0cm,rightmargin=0cm,%
roundcorner=5pt,
topline=false,bottomline=true,rightline=false,leftline=false,%
backgroundcolor=gray05,
frametitlebackgroundcolor=steBoxLine
}

\mdfdefinestyle{stebox5}{%
linecolor=steBoxLine,
linewidth=2pt,%
leftmargin=0cm,rightmargin=0cm,%
roundcorner=5pt,
topline=false,bottomline=false,rightline=false,leftline=true,%
backgroundcolor=gray05
}

\newcommand{\steResearchQuestionBox}[1]
{   
    \medskip
    \begin{mdframed}[style=stebox5]
    {#1}
    \end{mdframed}
    \medskip
}

\newcommand{\steSummaryBox}[2]
{   
    \medskip
    \begin{mdframed}[style=stebox4,frametitle={\textcolor{white}{\textbf{#1}}}]
    {#2}
    \end{mdframed}
    \medskip
}

\begin{document}

\title{Socio-Technical Well-Being of Quantum Software Communities: An Overview on Community Smells}

\titlerunning{Socio-Technical Well-Being of Quantum Software Communities}

\author{Stefano Lambiase\orcidID{0000-0002-9933-6203} \and
Manuel De Stefano\orcidID{0000-0001-6038-4171} \and
Fabio Palomba\orcidID{0000-0001-9337-5116} \and
Filomena Ferrucci\orcidID{0000-0002-0975-8972} \and
Andrea De Lucia\orcidID{0000-0002-4238-1425}}
\authorrunning{S. Lambiase et al.}
\institute{University of Salerno, Fisciano (SA), Italy
\email{\{slambiase,madestefano,fpalomba,fferrucci,adelucia\}@unisa.it}}

\maketitle

\begin{abstract}
Quantum computing has gained significant attention due to its potential to solve computational problems beyond the capabilities of classical computers. With major corporations and academic institutions investing in quantum hardware and software, there has been a rise in the development of quantum-enabled systems, particularly within open-source communities. However, despite the promising nature of quantum technologies, these communities face critical socio-technical challenges, including the emergence of socio-technical anti-patterns known as community smells. These anti-patterns, prevalent in open-source environments, have the potential to negatively impact both product quality and community health by introducing technical debt and amplifying architectural and code smells. Despite the importance of these socio-technical factors, there remains a scarcity of research investigating their influence within quantum open-source communities. This work aims to address this gap by providing a first step in analyzing the socio-technical well-being of quantum communities through a cross-sectional study. By understanding the socio-technical dynamics at play, it is expected that foundational knowledge can be established to mitigate the risks associated with community smells and ensure the long-term sustainability of open-source quantum initiatives.
\end{abstract}

\keywords{Quantum Software Enginering  \and Socio-Technical Aspects \and Community Smells \and Open Source Communities.}

%%%%%%%%%%%%%%%%%%%%%%%%%%%%%%%%%%%%%%%%%%%%%%%%%%%%%%%%%%%%%%%%%%%%%%%%%%%%%%%%%%%%%%%%%%
%%%%%%%%%%%%%%%%%%%%%%%%%%%%%%%%%%%%%%%%%%%%%%%%%%%%%%%%%%%%%%%%%%%%%%%%%%%%%%%%%%%%%%%%%%
\section{Introduction}

In recent years, \textit{quantum computing}, a field of computer science based on quantum theory, has gained significant attention in both research and industry. Quantum software technologies are increasingly adopted for their potential to solve computational problems beyond the capabilities of classical computers~\cite{knight2018serious, hoare2005grand}. As a result, major companies like IBM and Google have invested heavily in quantum hardware, offering users access to resources for experimentation and development. This has enabled the creation of \textit{quantum-enabled systems} that integrate quantum software into their operations, offering significant benefits for software development. For example, the exponential increase in processing power from quantum computing~\cite{zhang2020_recent_advances_in_quantum} could lead to higher-quality software products, while \textit{quantum machine learning (QML)} supports deeper analysis of large, complex datasets, advancing both machine learning and scientific research.

To democratize access to this transformative technology, the field of \textit{quantum software engineering (QSE)} has emerged~\cite{piattini2021toward, piattini2020talavera, piattini2021quantum, moguel2020roadmap}. Since the publication of the Talavera Manifesto, researchers have been actively designing and implementing advanced quantum software applications that leverage the computational capabilities of quantum computers~\cite{yarkoni2022quantum}. In parallel, significant efforts have been made to keep quantum software development largely an open-source practice~\cite{unitaryfund_survey, qosf_github, qosf_website, shaydulin2020making}. This approach is reinforced by academic studies that explore this phenomenon from different perspectives~\cite{shaydulin2020making, destefanoEmpiricalStudyCurrent}. An important observation is that today, it can be observed that the majority of the community involved in quantum software development operates within open-source contexts, as evidenced by the numerous repositories and contributors across various collaboration platforms~\cite{unitaryfund_survey, qosf_github, qosf_website, shaydulin2020making}.

However, the advantages of open-source activities come with significant socio-technical challenges. Extensive research has examined the socio-technical dynamics within open-source communities, resulting in the identification and classification of various socio-technical anti-patterns, which highlight problems in the organizational and collaborative structures of these  communities~\cite{social_debt, caballero2023community, tamburriExploringCommunitySmells2021, palombaTechnicalAspectsHow2021}. These issues can have serious consequences, potentially leading to critical project failures and, in the worst scenarios, the death of the community. Indeed, software development, by its nature, is a socio-technical activity~\cite{MyticalManMonth, Social_Theory, hoda2021_STGT}, involving stakeholders from diverse backgrounds who collaborate on innovative technologies. Poor management of this diversity can lead to subtle, often unnoticed social problems, eventually contributing to what is termed \textit{social debt}, the hidden costs of maintaining a development community with suboptimal dynamics~\cite{social_debt}. Researchers' work on identifying and addressing these problems has led to the concept of \textit{community smells}, socio-technical anti-patterns (e.g., excessive formality) and behavioral patterns (e.g., repeated condescension or sudden departures) that can exacerbate social debt~\cite{community_smell_def, community_smell_def_2}. These anti-patterns should not be underestimated. Research has demonstrated that community smells, particularly in open-source environments, have the potential to significantly harm both the community and the product. For instance, they can negatively impact product quality and introduce technical debt through the amplification of both architectural~\cite{tamburri2019software} and code smells~\cite{social_debt, palombaTechnicalAspectsHow2021}.  

Given that quantum software development is primarily conducted as an open-source activity, it can be argued that quantum communities are not immune to socio-technical anti-patterns like community smells. As a result, these communities may face socio-technical challenges that could impede their progress, which is particularly crucial during this pivotal phase of technological innovation that quantum computing and development are experiencing. Despite the relevance of these concerns, also reported by relevant literature~\cite{de2024quantum, shaydulin2020making}, \textbf{studies specifically investigating the socio-technical dimensions of quantum open-source communities are entirely absent.} This lack of research is problematic for several reasons. First, without socio-technical insights, it becomes difficult to identify and address challenges that could slow the development and adoption of quantum technologies, which are heavily dependent on collaborative efficiency. Second, the absence of research in this area limits the ability of community leaders to foster inclusive and sustainable environments, potentially leading to the exclusion of diverse talents that are vital for innovation. Finally, as quantum technologies continue to grow rapidly, unresolved community smells risk accumulating into social debt, which could increase long-term project costs and hinder the scalability of open-source quantum initiatives.

%\topar{Obiettivo del lavoro, fare un primo passo verso l'addressare questi fattori.}
To address the limitations mentioned above, we aimed to provide a first step into the investigation of the socio-technical well-being of quantum software communities by means of a statistical overview. By doing so, we aimed to depict the current situation of open-source communities developing quantum-enabled projects to put foundational knowledge and reasons for ulterior works in such a context. It is reasonable that this knowledge is essential: as highlighted earlier, the presence of socio-technical anti-patterns (like community smells) has been correlated with, and shown to influence, the emergence of other issues not only of a social nature but also strictly technological (such as architectural and code smells~\cite{tamburri2019software, social_debt, palombaTechnicalAspectsHow2021}). These issues have the potential to undermine a software community, significantly limiting, and in some cases completely nullifying its impact.

To achieve our goal, we carried out a \textit{cross-sectional study}~\cite{gordisEpidemiologyEBook2013, tillmanLearningCausalStructure2014}, which is a type of research designed to assess and illustrate essential traits of a population at a particular moment in time. Such studies offer a snapshot of the occurrence of a disease or condition (community smells) and the spread of various factors among a population (communities developing quantum-enabled systems).

%%%%%%%%%%%%%%%%%%%%%%%%%%%%%%%%%%%%%%%%%%%%%%%%%%%%%%%%%%%%%%%%%%%%%%%%%%%%%%%%%%%%%%%%%%
%%%%%%%%%%%%%%%%%%%%%%%%%%%%%%%%%%%%%%%%%%%%%%%%%%%%%%%%%%%%%%%%%%%%%%%%%%%%%%%%%%%%%%%%%%
\section{Background and Related Work}

This section describes the background and related work that is the foundation for our contributions.

%%%%%%%%%%%%%%%%%%%%%%%%%%%%%%%%%%%%
\subsection{Quantum Computing and Quantum Software Communities}

Quantum computing is a field within computer science that leverages the principles of quantum theory, specifically applying quantum mechanics to perform computations~\cite{knight2018serious, hoare2005grand}. Unlike classical computing, which relies on bits that take binary values (zero or one), quantum computing uses qubits, which can exist in a superposition of both zero and one states simultaneously. Quantum gates, which perform unitary transformations, are used to manipulate quantum information. Quantum programming languages manage both classical and quantum data using registers, and quantum programs are represented as quantum circuits, where gates are applied in a specific sequence. These circuits are executed on either real quantum hardware or simulators, with the results measured and stored in classical registers.

Quantum Software Engineering (QSE) is an emerging research field that has been formalized through the \textit{Talavera Manifesto}~\cite{piattini2020talavera}, which established the foundational principles for this discipline. In extending this contribution, Piattini et al.~\cite{piattini2021toward} focused on defining specific research domains within QSE. They identified four areas: the design of quantum-hybrid systems, testing methodologies, assessing the quality of quantum programs, and re-engineering classical-quantum information systems. A key point raised in their work is the need to bridge the knowledge gap between quantum computer scientists and traditional software engineers, as collaboration between these fields is essential for advancing QSE.

Building on this work, De Stefano et al.~\cite{destefanoQuantumFrontierSoftware2023} conducted a systematic mapping study to reveal the current state of research in QSE. Their study~\cite{destefanoQuantumFrontierSoftware2023} found that the primary focus of QSE research has been on software testing, highlighting a concentration of efforts in this area. However, their findings also suggest the need for a more balanced distribution of research efforts across other QSE domains. In particular, socio-technical aspects within quantum software development communities and software engineering management have received less attention, despite growing recognition of their importance. Both contributors~\cite{unitaryfund_survey} and researchers~\cite{de2024quantum, shaydulin2020making} have called for increased research in these areas to address the unique challenges posed by the interdisciplinary nature of quantum software development.

%%%%%%%%%%%%%%%%%%%%%%%%%%%%%%%%%%%%
\subsection{Socio-Technical Well-Being—Community Smells}

Software development and its engineering are inherently socio-technical activities. To assess the influence of social dynamics on software development, researchers—drawing on the well-established notion of Technical Debt\cite{palombaPredictingEmergenceCommunity2021, palombaTechnicalAspectsHow2021}—introduced the concept of \textit{Social Debt}, which refers to the unforeseen costs associated with sub-optimal decisions in collaboration, communication, and team management~\cite{caballero2023community, social_debt}. Additionally, in an effort to further characterize and identify the sources of Social Debt, researchers proposed the concept of \textit{Community Smell}, defined as socio-technical anti-patterns that may negatively impact the socio-technical well-being of a software development team, potentially leading to the accumulation of Social Debt~\cite{caballero2023community}.

The research explored the relationship between community smells and various aspects of software development. Notably, Palomba et al.~\cite{palombaTechnicalAspectsHow2021} examined the connection between community smells and code smells, their product-oriented counterpart, demonstrating that community smells are among the primary factors influencing the emergence of code smells. Tamburri et al.~\cite{tamburriExploringCommunitySmells2021} conducted a large-scale study across 60 open-source ecosystems to assess (1) the diffusion of community smells and (2) their perceived impact by developers, revealing that community smells are both widespread and perceived to affect the evolution and sustainability of software communities. These findings indicate that socio-technical antipatterns can influence maintenance and evolution in two ways: by directly increasing social debt (i.e., increasing costs related to socio-technical problems) and by affecting product-related factors, thereby increasing technical debt. Furthermore, two mining studies conducted by Catolino et al.~\cite{catolino2019gender} and Lambiase et al.~\cite{lambiaseGoodFencesMake2022} revealed correlations between the emergence of community smells and gender diversity (in the former) and cultural heterogeneity (in the latter). 

Regarding detection methods, Palomba and Tamburri~\cite{codeface4smells} proposed a machine learning approach for predicting community smells based on socio-technical metrics, achieving promising results with an F-measure of 78\%. Additionally, Almarimi et al.~\cite{almarimiLearningDetectCommunity2020} introduced a multi-label learning model using genetic algorithms to detect ten community smells and developed the community smells detection tool \textsc{csDetector}~\cite{almarimiCsDetectorOpenSource2021}. Building on Almarimi et al.'s work~\cite{almarimiCsDetectorOpenSource2021, almarimiLearningDetectCommunity2020}, Voria et al.~\cite{voria2022_CADOCS} developed \textsc{CADOCS}, a conversational agent capable of detecting ten community smells from a software repository and proposing refactoring strategies for some of them. Moreover, Advanced network models, such as the MOGen higher-order network model, have been developed to detect community smells by analyzing complex relationships and interaction patterns within software teams~\cite{gote2023locating}.

\steSummaryBox{\faList\ Related Work: Summary and Research Gap.}{The effort to keep quantum computing tied to open source has not been matched by efforts to understand the socio-technical challenges of these communities. This gap limits contributors' ability to assess community health and restricts QSE researchers from exploring a well-established area of software engineering.}

%%%%%%%%%%%%%%%%%%%%%%%%%%%%%%%%%%%%%%%%%%%%%%%%%%%%%%%%%%%%%%%%%%%%%%%%%%%%%%%%%%%%%%%%%%
%%%%%%%%%%%%%%%%%%%%%%%%%%%%%%%%%%%%%%%%%%%%%%%%%%%%%%%%%%%%%%%%%%%%%%%%%%%%%%%%%%%%%%%%%%
\section{Cross-sectional Study—An Overview}

This section provides the reader with basic knowledge of observational studies~\cite{wangCrossSectionalStudiesStrengths2020, tillmanLearningCausalStructure2014}. For space limitations, we suggest reading the work of Saarimäki et al.~\cite{saarimakiCohortStudiesSoftware2020, saarimakiRobustApproachAnalyze2022} to gain a better understanding of the method.

\subsection{Cross-Sectional Studies}

\textit{Observational studies}, including \textit{cross-sectional studies}, are widely used in epidemiology to examine associations between exposures and outcomes without intervention~\cite{wangCrossSectionalStudiesStrengths2020, tillmanLearningCausalStructure2014}. The main types include \textit{cohort}, \textit{case-control}, and \textit{cross-sectional} designs~\cite{tillmanLearningCausalStructure2014, wangCrossSectionalStudiesStrengths2020}.

\textit{Cross-sectional studies} capture population characteristics at a single time point, providing a snapshot of prevalence and exposure distribution~\cite{wangCrossSectionalStudiesStrengths2020}. Data are collected simultaneously from all participants, enabling efficient assessment of existing conditions. \textit{Prevalence}—the proportion of individuals with a given condition—is central to these studies~\cite{wangCrossSectionalStudiesStrengths2020}. The \textit{Prevalence Odds Ratio (POR)} quantifies the association between an exposure and a condition by comparing the odds of exposure in affected versus unaffected individuals~\cite{tillmanLearningCausalStructure2014, wangCrossSectionalStudiesStrengths2020}.

Cross-sectional studies are time- and cost-efficient and useful for estimating prevalence and generating hypotheses~\cite{tillmanLearningCausalStructure2014, wangCrossSectionalStudiesStrengths2020}. However, they cannot determine causality or temporal order and may be affected by biases from self-reported data.

%%%%%%%%%%%%%%%%%%%%%%%%%%%%%%%%%%%%%%%%%%%%%%%%%%%%%%%%%%%%%%%%
\begin{table*}
    \renewcommand{\arraystretch}{1.25}
    %\scriptsize
    \centering % instead of \begin{center}
    \caption{Community Smells investigated in our study.}

    \rowcolors{1}{graytable}{white}
    \begin{tabularx}{1\linewidth}{l X}
        \rowcolor{black}
        \multicolumn{1}{l}{\textcolor{white}{\textbf{Community Smell}}} & \multicolumn{1}{X}{\textcolor{white}{\textbf{Definition}}} \\
        Organizational Silo (OSE) & Siloed areas of the community that do not communicate, except through one or two of their members.\\
        Black Cloud (BCE) & Information overload due to a lack of structured communications or cooperation governance.\\
        Radio Silence (RS) & One interposes herself into every formal interaction across more sub-communities with little flexibility to introduce other channels.\\
        Prima Donnas (PDE) & A team member is unwilling to respect external changes from other team members.\\
        Sharing Villainy (SV) & Cause of a lack of information exchange, team members share essential knowledge such as outdated, wrong, and unconfirmed information.\\
        Organizational Skirmish (OS) & A misalignment between different expertise levels of individuals involved in the project leads to dropped productivity and affects the project's timeline and cost.\\
        Solution Defiance (SD) & The development community presents different levels of cultural and experience background, leading to the division of the community into similar subgroups with completely conflicting opinions.\\
        Truck Factor Smell (TF) & Risk of significant knowledge loss due to the turnover of developers resulting from the fact that project information and knowledge are concentrated in a minority of the developers.\\
        Unhealthy Interaction (UI) & Long delays in stakeholder communications cause slow, light and brief conversations and discussions.\\
        Toxic Communication (TC) & Communications between developers are subject to toxic conversations and negative sentiments containing unpleasant, anger or even conflicting opinions towards various issues that people discuss.\\
        \hline
    \end{tabularx}
    \label{table:community_smells}
\end{table*}

\subsection{Mining Software Repositories as Observational Studies}

The Software Engineering research community has seen a significant increase in studies that use mining software repositories (MSR), with the rise in popularity of online code repository platforms like GitHub. 
This has led to the introduction of rules of thumb, highlighting common pitfalls~\cite{kalliamvakouPromisesPerilsMining2014} to improve the quality of these studies and their outcomes.

However, it is important to note that these studies cannot provide causal explanations for observed phenomena, despite their usefulness and straightforward execution. 
To address this limitation, Saarimäki et al.~\cite{saarimakiCohortStudiesSoftware2020, saarimakiRobustApproachAnalyze2022} recommended the adoption of observational studies, particularly cohort studies, which offer the highest level of scientific evidence.

In the context of QSE, a relatively emerging discipline with limited guidelines and tools for investigating socio-technical issues, a cross-sectional study is justified for several reasons. 
Firstly, guidelines for MSR studies are somewhat limited~\cite{saarimakiCohortStudiesSoftware2020}, and pitfalls abound~\cite{kalliamvakouPromisesPerilsMining2014}. 
Secondly, in the field of QSE, socio-technical aspects have not been extensively explored~\cite{destefanoQuantumFrontierSoftware2023}.
A cross-sectional study hence represents a pragmatic and cost-effective initial step to spark the investigation of socio-technical aspects in QSE, which in this context are represented by community smells and the relationships among them. 

Research on socio-technical aspects of QSE is a new area that has not been explored yet~\cite{destefanoQuantumFrontierSoftware2023}. 
Cross-sectional studies can provide valuable insights into this field's current conditions and associations. 
These studies are especially helpful in generating hypotheses and exploring potential connections. 
Despite the limitations, cross-sectional studies provide a practical starting point for further research.

%%%%%%%%%%%%%%%%%%%%%%%%%%%%%%%%%%%%%%%%%%%%%%%%%%%%%%%%%%%%%%%%%%%%%%%%%%%%%%%%%%%%%%%%%%
%%%%%%%%%%%%%%%%%%%%%%%%%%%%%%%%%%%%%%%%%%%%%%%%%%%%%%%%%%%%%%%%%%%%%%%%%%%%%%%%%%%%%%%%%%

\section{Research Design}\label{sec:experiment}

%\topar{Obiettivo}
The \textit{goal} of this research was to examine the socio-technical well-being of open-source software communities developing quantum software enabled-systems. The \textit{purpose} was to uncover foundational knowledge able to (1) inform open-source contributors' future choices and (2) shed light on future research agenda on socio-technical aspects in the field of QSE.

%\topar{Domande di ricerca}
In order to reach our objective, we operationalized the socio-technical well-being of open-source communities using community smells. At first, we wanted to understand the current situation of such communities, aiming to depict the current diffusion of community smells. Identifying how widespread community smells are in quantum computing projects helps to highlight socio-technical challenges that can impact the productivity and health of these open-source communities. Thus, we formulated the following research question:

\steResearchQuestionBox{\faQuestionCircleO \hspace{0.05cm} \textbf{RQ\textsubscript{1}}: \textit{What is the prevalence of community smells in quantum projects?}}

After assessing the diffusion of smells, in order to better characterize the socio-technical well-being of open-source quantum software communities and the phenomenon of community smells inside them, we investigated the correlation between the different smells in the same community. Exploring relationships between different community smells can reveal deeper socio-technical issues, helping us understand how these problems interact and impact community dynamics. Thus, we formulated the following second research question: 

\steResearchQuestionBox{\faQuestionCircleO \hspace{0.05cm} \textbf{RQ\textsubscript{2}}: \textit{Is there any relationship between different community smells in the context of quantum projects?}}

%\topar{Metodo di ricerca}
To answer the research questions, we conducted a cross-sectional study. First, we selected a set of quantum-enabled software projects on GitHub and extracted community smells from them. Then, to answer RQ\textsubscript{1}, we computed the \textit{prevalence} of the community smells, while, to answer RQ\textsubscript{2}, we computed the \textit{Prevalence Odds Ratio} to assess the correlations between community smells. Further details about our research process are in the following sections and in our online appendix~\cite{online_appendix}.

%%%%%%%%%%%%%%%%%%%%%%%%%%%%%%%%%%%%
\begin{table}
    \renewcommand{\arraystretch}{1.25}
    \centering % instead of \begin{center}
    \caption{Metrics for the 17 analyzed repositories.}

    \rowcolors{1}{graytable}{white}
    \begin{tabularx}{1\linewidth}{X llll}
        \hline
        \rowcolor{black}
        \multicolumn{1}{X}{\textcolor{white}{\textbf{Repository}}} & 
        \multicolumn{1}{l}{\textcolor{white}{\textbf{Contributors}}} & 
        \multicolumn{1}{l}{\textcolor{white}{\textbf{Commits}}} & 
        \multicolumn{1}{p{0.08\columnwidth}}{\textcolor{white}{\textbf{Stars}}} & 
        \multicolumn{1}{p{0.15\columnwidth}}{\textcolor{white}{\textbf{Start Date}}}\\
        \hline
        qrand & 3 & 287 & 22 & 2020-10-14 \\
        bloch\_sphere & 3 & 27 & 76 & 2020-06-01 \\
        tweedledum & 6 & 263 & 86 & 2018-07-13 \\
        xacc & 21 & 2546 & 138 & 2017-09-19 \\
        tequila & 35 & 1313 & 305 & 2020-04-28 \\
        qsearch & 3 & 895 & 29 & 2019-05-29 \\
        quantpy & 3 & 20 & 13 & 2017-09-28 \\
        QTensor & 6 & 453 & 40 & 2018-07-23 \\
        quantum\_decomp & 2 & 77 & 21 & 2019-05-06 \\
        node-red-contrib-quantum & 7 & 148 & 13 & 2021-06-16 \\
        nanite & 11 & 667 & 14 & 2012-01-24 \\
        scikit-quant & 3 & 220 & 34 & 2019-01-10 \\
        dc-qiskit-qml & 2 & 85 & 10 & 2019-01-13 \\
        quantum-robot & 3 & 199 & 4 & 2020-06-22 \\
        OpenFermion-FQE & 10 & 404 & 42 & 2020-04-01 \\
        shor & 4 & 80 & 8 & 2020-02-19 \\
        OpenFermion-Cirq & 15 & 254 & 267 & 2018-03-20 \\
        \hline
    \end{tabularx}
    \label{tab:repo_metrics}
\end{table}

\subsection{Population and Data Collection}

To answer our research questions, we investigated the contributors of communities that actively participate in the development of open-source quantum-related software. Thus, we chose to focus on a representative subset of the target population due to logistical limitations in collecting information across all quantum open-source projects on GitHub.

We took advantage of a previous dataset of 115 repositories published in a paper by De Stefano et al.~\cite{destefanoSoftwareEngineeringQuantum2022a}. The number of contributors per repository ranges from just 1 to a maximum of 10. This indicates that most repositories have a small core team, while a few involve slightly larger groups. The number of commits varies widely, with a median value of 65, indicating significant differences in development activities across repositories. The number of stars, used as an indicator of popularity, also shows considerable variability, with a median of 10 stars per repository, suggesting a generally modest level of community recognition.

After selecting the communities for analysis, community smells were computed for each one project. The augmented version of \textsc{csDetector}~\cite{almarimiCsDetectorOpenSource2021},\footnote{\textsc{csDetector} augmented: \url{https://github.com/gianwario/csDetector}} available in the main repository of \textsc{CADOCS}~\cite{voria2022_CADOCS}, was used for this task. It is important to note that the tool can detect ten types of community smells, reported in Table \ref{table:community_smells}. Specifically, the tool was applied with its default parameters to each repository in the selected dataset of quantum open-source repositories. To work properly, \textsc{csDetector} requires the URL of the repository and that certain criteria are met.\footnote{\textsc{csDetector} criteria include (1) the presence of commits, (2) the existence of multiple authors, (3) the availability of a main branch, (4) at least one pull request (or more), (5) ensuring that the messages associated with these pull requests are not empty, and (6) at least one issues (or more).}

Ultimately, 17 repositories (described in Table \ref{tab:repo_metrics}) out of the original 115 were successfully analyzed.\footnote{ It is important to note that various state-of-the-art tools were tested to detect community smells~\cite{codeface4smells, paradis2024analyzing}, but \textsc{csDetector} was the only one capable of analyzing a subset of the original dataset.} These repositories have varying numbers of contributors, ranging from 2 to 35, reflecting different scales of community involvement. The number of commits ranges from 20 to 2546, indicating diverse levels of development activity. The repositories also have different levels of community recognition, as shown by their stars, which range from 4 to 305. The opening and last commit dates provide insights into the lifecycle of each repository, showcasing both long-standing and more recent projects.

%%%%%%%%%%%%%%%%%%%%%%%%%%%%%%%%%%%%
\subsection{Data Analysis}

To determine the prevalence of each community smell (thus, to answer the first research question), we computed the prevalence denoted as \(P(X)\)~\cite{wangCrossSectionalStudiesStrengths2020}, which represents the ratio of repositories where a specific smell was identified to the total number of repositories analyzed: \(P(X) = \frac{\text{Repositories with Smell X}}{\text{Total Repositories}}\)

\begin{comment}
\begin{wraptable}{r}{0.5\linewidth}
    \centering % instead of \begin{center}
    \vspace{-10pt}
    \caption{POR Contingency Table.}
     
    \resizebox{\linewidth}{!}{
   
    \begin{tabular}{ccc}
        \toprule
         & \textbf{A Present} & \textbf{A Absent} \\
        \midrule
        \textbf{B Present} & A & B \\
        \textbf{B Absent} & C & D \\
        \bottomrule
        \end{tabular}
    }
    \vspace{-10pt}
    \label{table:contingency_table}
\end{wraptable}
\end{comment}

To assess correlations among community smells (thus, to answer RQ\textsubscript{2}), we employed the POR as a statistical tool~\cite{wangCrossSectionalStudiesStrengths2020}. It evaluates how the presence of one condition ($X_1$) correlates with the presence or absence of another condition ($X_2$)—community smells, in our context. The formula for POR is \(POR = \frac{AD}{BC}\)~\cite{wangCrossSectionalStudiesStrengths2020}, where: \(A\) is the number of cases where both conditions ($X_1$ and $X_2$) are present, \(B\) is the number of cases where condition $X_1$ is present, but condition $X_2$ is absent, \(C\) is the number of cases where condition $X_1$ is absent, but condition $X_2$ is present, and \(D\) is the number of cases where both conditions ($X_1$ and $X_2$) are absent.

\begin{comment}
The POR quantifies the strength of the correlation between an exposure factor and a case (community smells, in our context). 
Given the Contingency Table \ref{table:contingency_table}, in which a specific community smell (B) is the exposure and another community smell (A) is the case, we computed the POR using the formula \(POR = \frac{AD}{BC}\)~\cite{wangCrossSectionalStudiesStrengths2020}.
\end{comment}

To further interpret the POR results, values significantly above 1 suggest a strong positive association between the community smells. For example, a POR of 2 would indicate that community smell $X_1$ is twice as likely to occur in communities where community smell $X_2$ is present, compared to those where it is absent. On the other hand, a POR less than 1 indicates a negative correlation, meaning the presence of smell $X_2$ decreases the likelihood of smell $X_1$. For instance, a POR of 0.5 would imply that community smell $X_1$ is half as likely to occur in the presence of smell $X_2$, suggesting that smell B may play a mitigating role.

%%%%%%%%%%%%%%%%%%%%%%%%%%%%%%%%%%%%
\subsection{Threat to Validity}

The study acknowledges the presence of threats to validity that could impact the integrity of the findings.

Regarding \textit{construct validity}, the study's scope is limited to a subset of known community smells due to tool limitations, which could potentially overlook certain socio-technical challenges within quantum developer communities. 
However, despite this limitation, the study highlights the importance of considering community smells in software development. 
It is worth noting that every component in a repository may be affected by community smells, and while the assumption made in the study may not be entirely accurate, it serves as a useful starting point. 
Furthermore, the technical limitations of the employed tools should be taken into account when interpreting the study's results. 
Overall, this study provides valuable insights into the impact of community smells on software development while also highlighting the need for further research in this area.

Concerning \textit{conclusion validity}, the cross-sectional nature of the study design poses a threat to drawing causal conclusions between exposure factors (community smells) and outcomes. 
The absence of longitudinal data precludes establishing temporal relationships or causal links.

Regarding \textit{internal validity}, a potential selection bias arises from the reliance on repositories analyzed in a previous study, 
limiting the generalizability of results to the entire quantum open-source landscape. 
Additionally, the application of the \textsc{csDetector} tool introduces the possibility of measurement bias, as not all community smells may be detected.

Concerning \textit{external validity}, the study's findings cannot be universally generalized to all quantum open-source repositories, as the sample was selected based on a previous study and analyzed using a specific tool. 
Therefore, the broader quantum developer community may not be fully represented.

%%%%%%%%%%%%%%%%%%%%%%%%%%%%%%%%%%%%%%%%%%%%%%%%%%%%%%%%%%%%%%%%%%%%%%%%%%%%%%%%%%%%%%%%%%
%%%%%%%%%%%%%%%%%%%%%%%%%%%%%%%%%%%%%%%%%%%%%%%%%%%%%%%%%%%%%%%%%%%%%%%%%%%%%%%%%%%%%%%%%%
\begin{figure}[h]
    \centering    
    \includegraphics[width=\columnwidth]{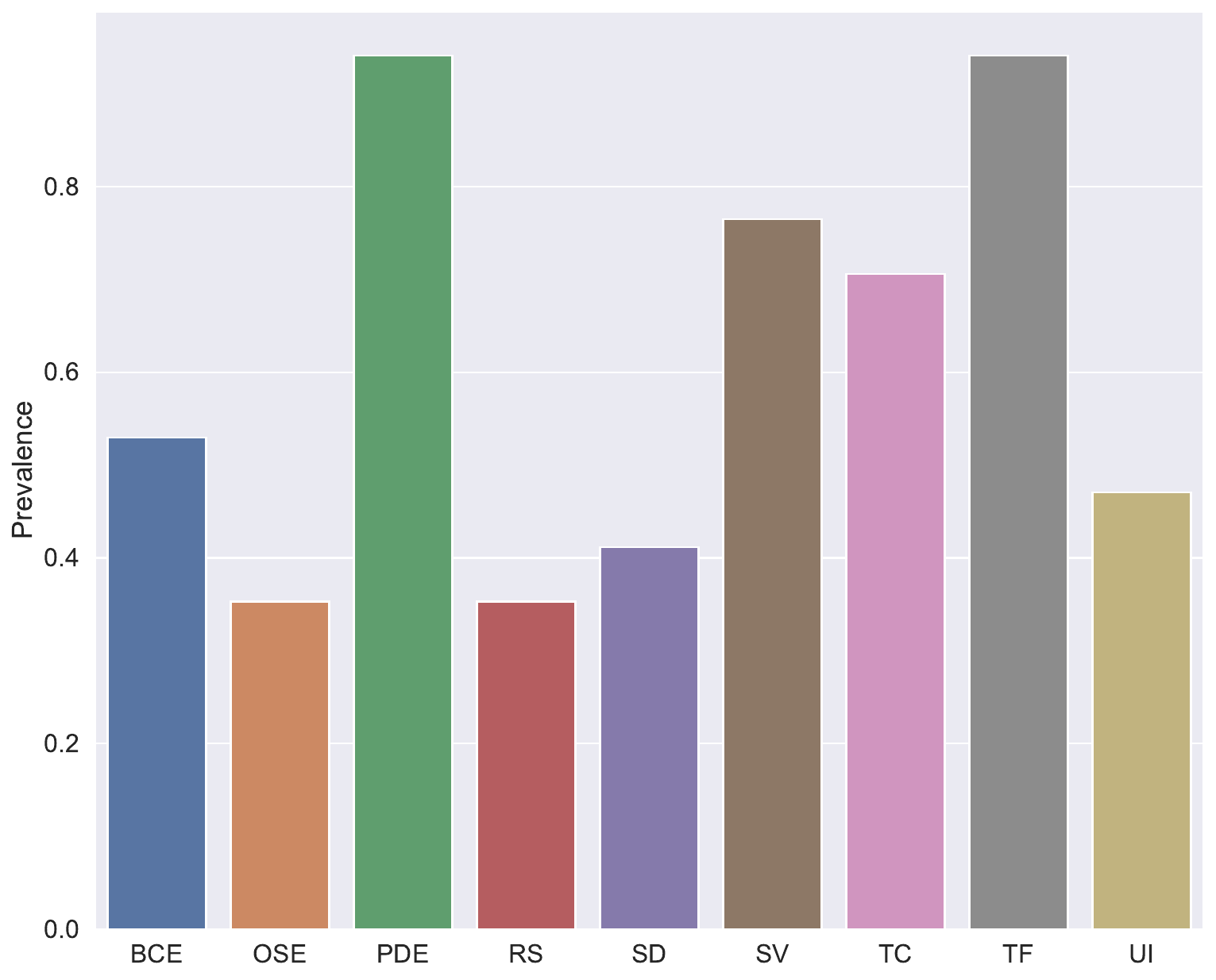}
    \caption{Prevalence of community smells affecting the sampled repositories.}
    \label{fig:community_smell_prevalence}
\end{figure}

\section{Results}\label{sec:results}

In this section, we present the results of our analysis that we conducted and described in \Cref{sec:experiment}. 

\Cref{fig:community_smell_prevalence} depicts the prevalence of community smells that we found in the sample.
What is most evident is that more than half of the considered smells—i.e., Black Cloud Effect (BCE), Power Distance Effect (PDE), Sharing Villainy (SV), Toxic Communication (TC), and Truck Factor (TF)—have a prevalence of more than 50\%. 
This implies that these particular smells are pervasive within the analyzed repositories, with more than half of the repositories exhibiting them.
The most extreme values are shown by PDE and TF, which occur in 94\% of the cases.
On the flipped side, Radio Silence (RS) and Organizational Silo Effect (OSE) are the least occurring smells, both exhibiting a prevalence of 35\%, while Solution Defiance (SD) exhibits a prevalence of 40\%.
What is also interesting is that Organizational Skirmish (OS) is the only smell that never occurs in any of the considered repositories. 

These findings provide a direct answer to RQ\textsubscript{1}, showing that several community smells—most notably PDE and TF—are highly prevalent in quantum projects, with over half of the repositories affected by at least five different smells.

\begin{figure}[h]
    \centering    
    \includegraphics[width=\columnwidth]{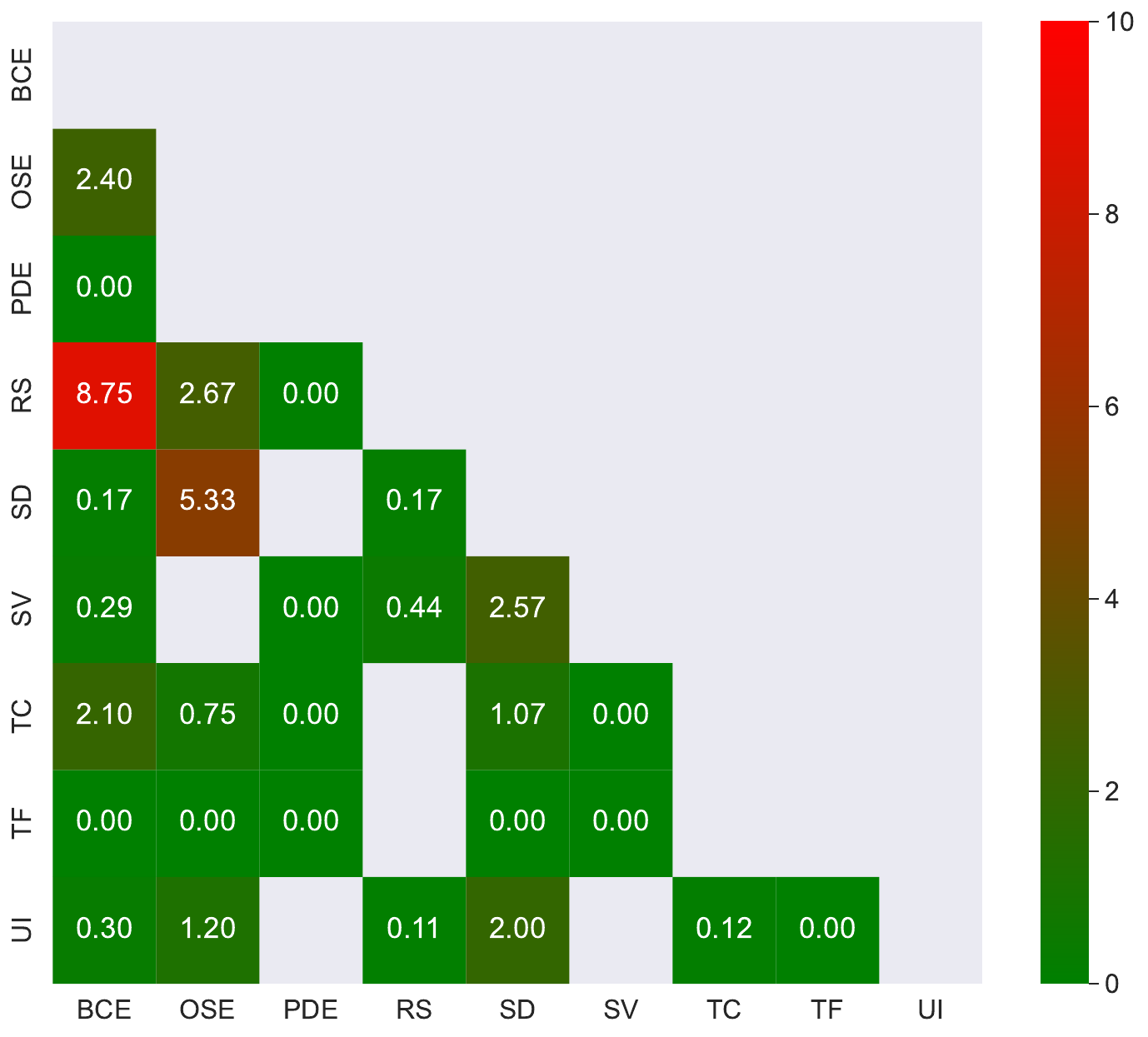}
    \caption{Prevalence Odds Ratio among the community smells affecting the sampled repositories.}
    \label{fig:community_por}
\end{figure}

In \Cref{fig:community_por}, we can see the matrix of the POR that affects the considered repositories.
What can be immediately seen is that there is no smell that has only positive correlations with the other smells.
On the contrary, there are some smells that have only negative correlations.
Starting from the positive correlations, we discovered some interesting patterns from our analysis. 
For instance, if RS is present in a repository, it is highly likely to encounter BCE, as they have a strong association (POR 8.750). 
Similarly, if a repository has SD, it is more likely to feature OSE, as they have an association (POR 5.333). 
We also found moderate positive correlations between several other community smells: RS and OSE have a POR of 2.667, SV and SD have a POR of 2.571, OSE and BCE have a POR of 2.400, and TC and BCE have a POR of 2.100. 
The presence of UI moderately associates with the presence of SD (POR 2.000) but only weakly associates with OSE (POR 1.200). 
Lastly, we observed no correlation between TC and SD (POR 1.07).

The observations where the POR is below 1 clearly indicate a significantly lower likelihood of co-occurrence.
For example, we noted a POR of 0.75 which indicates a negative correlation between TC and OSE. 
This means that repositories containing TC are less likely to have OSE.
Similarly, when compared to RS, SV exhibited a POR of 0.444, which indicates a reduced likelihood of finding RS in repositories featuring SV. 
We also discovered additional negative correlations. 
Notably, SD and BCE exhibit a POR of 0.171, signifying a strong negative correlation. 
This indicates that when SD is present, repositories are significantly less likely to exhibit BCE. 
Similar negative correlations were identified between SD and RS, as well as UI and BCE, TC, and RS. 
Interestingly, a few pairs exhibited a POR of 0.0, suggesting a complete absence of correlation between these community smells. These pairs included various combinations, such as PDE and BCE, TF and PDE, TF and SV, TF and SD, TC and SV, TF and OSE, TF and BCE, RS and PDE, TC and PDE, SV and PDE, and UI and TF.

In summary, these findings address RQ\textsubscript{2} by confirming that community smells in quantum projects exhibit both strong positive and negative relationships, with some pairs frequently co-occurring and others rarely appearing together.

%\steSummaryBox{\faHandORight \hspace{0.05cm} RQ\textsubscript{2}: Relationships between community smells}{After analyzing community smells in quantum computing projects, both positive and negative correlations were found. A strong positive correlation (POR: 8.750) was observed between RS and BCE, while a negative correlation (POR: 0.75) was found between TC and OSE. Some other less strong positive and negative correlations were observed as well. Additionally, some pairs had a POR of 0.0, indicating no correlation.}

%%%%%%%%%%%%%%%%%%%%%%%%%%%%%%%%%%%%%%%%%%%%%%%%%%%%%%%%%%%%%%%%%%%%%%%%%%%%%%%%%%%%%%%%%%
%%%%%%%%%%%%%%%%%%%%%%%%%%%%%%%%%%%%%%%%%%%%%%%%%%%%%%%%%%%%%%%%%%%%%%%%%%%%%%%%%%%%%%%%%%
\section{Discussions and Implications}

This section reports some discussion points and implications that better contextualize our findings and could open up future work in the context of quantum software communities.

%%%%%%%%%%%%%%%%%%%%%%%%%%%%%%%%%%%%
\subsection{RQ\textsubscript{1}: On the Prevalence of Community Smells}

The results demonstrate that community smells, particularly Black Cloud (BCE), Prima Donnas (PDE), Sharing Villainy (SV), Toxic Communication (TC), and Truck Factor (TF), are highly prevalent within quantum software projects. With over half of the repositories exhibiting these smells, the data suggest that these issues are embedded in the fabric of quantum open-source communities. Notably, the extreme prevalence of PDE and TF, at 94\%, indicates structural problems in collaboration and knowledge retention. 

Compared to classical open-source communities, as explored by Tamburri et al.~\cite{tamburriExploringCommunitySmells2021} and reported by Caballero-Espinosa et al.~\cite{caballero2023community}, quantum repositories exhibit a distinct pattern of smell manifestation. While classical settings often report Bottleneck, Lone Wolf, and Organizational Silo, such smells rarely reach similar extremity. These differences likely stem from the relative immaturity of quantum ecosystems, which are shaped by niche expertise, smaller contributor bases, and tightly coupled development workflows. A cross-domain comparison with ML-enabled systems further supports this interpretation. Building on the work by Annunziata et al.~\cite{annunziata2025communities}, which shares the same detection strategy as ours, both PDE and TF again emerged as dominant, confirming that knowledge concentration and authority imbalance are recurring issues in high-tech domains. However, quantum repositories exhibited higher prevalence of TC, possibly reflecting the greater communicative friction introduced by the abstract and rapidly evolving nature of quantum computing. These patterns underscore the necessity of domain-aware socio-technical strategies when managing emerging software ecosystems.

Our findings have significant implications both for researchers and contributors to these projects:
\begin{itemize}
    \item For \textit{researchers}, the results highlight the need for targeted studies on the role of community smells—particularly PDE and TF—in shaping the sustainability of quantum software projects. These smells were not only highly prevalent but more extreme than in classical or ML-enabled systems, suggesting domain-specific coordination challenges. Future work should investigate whether such smells hinder innovation or community growth and explore interventions like improved governance or communication structures to mitigate them. This would extend socio-technical theory into the context of emerging, expertise-driven ecosystems.
    \item For \textit{open-source contributors}, the prevalence of PDE and TF indicates risks related to centralization and knowledge retention. While smaller teams and uniform expertise—reflected in the absence of OS—may ease collaboration, they also require proactive practices to prevent silos and ensure continuity. Lightweight mentoring, onboarding strategies, and documentation can help reduce these risks and support healthier project evolution.
\end{itemize}

\subsection{RQ\textsubscript{2}: On the Relationship Between Community Smells}

The results reveal a complex network of both positive and negative correlations between community smells in quantum software projects, providing valuable insights for both researchers and open-source contributors. The significant positive correlations, such as those between Radio Silence (RS) and Black Cloud (BCE) (POR 8.750), as well as Solution Defiance (SD) and Organizational Skirmish (OSE) (POR 5.333), suggest that these particular smells often co-occur. This may indicate systemic issues in communication and collaboration, which reinforce each other. For instance, the strong link between RS and BCE highlights the potential for communication breakdowns leading to information overload, while the connection between SD and OSE points to the misalignment of expertise exacerbating silos within the community.

On the other hand, the negative correlations, such as that between TC and OSE (POR 0.75), suggest that certain smells are less likely to appear together. This could imply that when one issue is present, it mitigates or prevents the formation of others, possibly due to counterbalancing effects in community dynamics. The negative correlation between SD and BCE (POR 0.171) supports this idea, suggesting that repositories struggling with conflicting opinions and expertise may paradoxically avoid issues related to unstructured communication.

Unlike classical open-source systems, for which—to the best of our knowledge—no existing studies analyze correlations between community smells, ML-enabled projects provide a useful comparison always in the work by Annunziata et al.~\cite{annunziata2025communities}. The co-occurrence patterns observed in quantum repositories—such as RS–BCE (POR 8.750) and SD–OSE (POR 5.333)—were notably stronger than those found in ML projects, suggesting tighter coupling between specific communication and coordination issues in the quantum domain. While ML repositories also exhibit elevated PORs for pairs like OSE–PDE or BCE–OSE, these associations tend to be more evenly distributed and of lower magnitude. Moreover, the greater number of smell pairs with POR equal to zero in quantum repositories—e.g., PDE–BCE, TF–PDE, and TC–SV—points to a more fragmented or modular manifestation of smells, in contrast to the more interconnected landscape of ML systems. Finally, negative associations were more frequent in quantum projects, indicating divergent team dynamics or counterbalancing patterns that are less apparent in ML contexts. These differences likely reflect the early-stage and specialized nature of quantum software development.

Also here we can provide some implications:
\begin{itemize}
    \item For \textit{researchers}, the results highlight the need to explore how specific smells interact and whether they trigger or reinforce one another. Strong associations—such as RS–BCE—may indicate cascading coordination failures, while the absence of overlap between smells like PDE–BCE suggests contextual or structural separation. Compared to ML-enabled systems, the sharper and more polarized correlations in quantum repositories call for domain-sensitive models that move beyond treating smells in isolation and toward understanding their interdependencies.

    \item For \textit{open-source contributors}, these patterns suggest that addressing certain high-risk combinations—such as RS and BCE—may reduce the impact of multiple smells simultaneously. The presence of negative correlations, such as TC–OSE, implies that mitigating one issue may lower the risk of others. Unlike the more entangled smell networks in ML projects, quantum repositories may benefit from focused interventions targeting the most disruptive pairs.
\end{itemize}

\section{Conclusion}

The findings of our study revealed not only the widespread presence of community smells but also significant correlations between different smells. As called for in the introduction, this research represents a preliminary yet foundational step toward a more comprehensive investigation of socio-technical issues in the QSE domain.
In terms of contributions, we provided: 
\begin{enumerate} 
    \item an empirical investigation of the presence of ten community smells in open-source quantum communities; 
    \item a statistical analysis of the correlations between pairs of these ten smells within the same communities; and 
    \item a publicly available online appendix~\cite{online_appendix}, provided to ensure replicability and reliability of the findings. 
\end{enumerate}

This study opens promising avenues for future research. A more detailed investigation into the specific socio-technical dynamics that lead to the emergence of these community smells is necessary. Such studies could employ qualitative methods to further explore the circumstances surrounding these smells and to identify effective mitigation strategies. Moreover, understanding these socio-technical issues can inform the technical side of software development. Previous research has suggested a link between community smells and technical debt, and future work should aim to confirm this relationship within the quantum software context. 

%%%%%%%%%%%%%%%%%%%%%%%%%%%%%%%%%%%%%%%%%%%%%%%%%%%%%%%%%%%%%%%%%%%%%%%%%%%%%%%%%%%%%%%%%%
%%%%%%%%%%%%%%%%%%%%%%%%%%%%%%%%%%%%%%%%%%%%%%%%%%%%%%%%%%%%%%%%%%%%%%%%%%%%%%%%%%%%%%%%%%
\begin{credits}
\subsubsection{\ackname} This work has been partially supported by the project ‘QUASAR: QUAntum software engineering for Secure, Affordable, and Reliable systems, grant 2022T2E39C, under the PRIN 2022 MUR program funded by the EU NextGenerationEU.

\subsubsection{\discintname}
The authors have no competing interests to declare that are relevant to the content of this article.
\end{credits}

\balance
\bibliographystyle{splncs04}
\bibliography{references}

\end{document}